\documentclass{svproc}
\usepackage{url}

\usepackage{graphics,graphicx}
\RequirePackage{doi}
\usepackage{newtxtext,newtxmath}
\hypersetup{%
	pdfborder = {0 0 0},
	colorlinks,
	citecolor=magenta,
	filecolor=Dark green,
	linkcolor=blue,
	urlcolor=blue
}
\usepackage[square,numbers]{natbib}

\begin{document}
\mainmatter              
\title{The Dark Matter Search at Jaduguda Underground Science Laboratory}
\titlerunning{DM Search at JUSL}  
%
\author{Vimal Kumar\inst{1,2} Susmita Das\inst{1,2}
Mala Das\inst{1,2*} for InDEX Collaboration}
\authorrunning{V. Kumar et al. (2025)} 
%
%
\institute{Saha Institute of Nuclear Physics, 1/AF, Salt Lake, Kolkata-700064, India\\
\and
Homi Bhabha National Institute, Training School Complex, Anushakti Nagar, Mumbai 400094, India \\
$^*$Corresponding Author:
\email{mala.das@saha.ac.in}}

\maketitle              

\begin{abstract}
The first run for the dark matter direct search experiment at Jaduguda Underground Science Laboratory is presented in this article. The experiment named InDEx; the Indian Dark matter search Experiment has been initiated with superheated emulsion detector consisting of the droplets of tetrafluoroethane ($C_2H_2F_4$). The detector ran for an effective period of 48.6 days at a threshold of 5.87 keV with an exposure of 2.47 kg-days. It is observed that the minimum sensitivity for the SI cross-section of $[7.939\pm(0.375)_{statistical}(^{+1.386}_{-0.909})_{systematic}]\times 10^{-39}$ $cm^2$ for fluorine appears at WIMP mass of 30.67 $GeV/c^2$. The InDEx with larger exposures is under development and yet to come in near future.

\end{abstract}
\section{Introduction}
The direct dark matter search experiments are mainly looking for the nuclear recoil signals arising from the WIMP-nucleus elastic scattering with different types of detector and detection techniques. The world leading direct detection experiments are mostly sensitive in 30-40 $GeV/c^2$ WIMP mass region and the null results from those experiments have pushed the interest to explore the low WIMP mass region specially below 20 $GeV/c^2$ \cite{2024arXiv241017036A,Schumann_2019}. To be sensitive in the lower mass region, the detector needs to be operated at very low threshold which is a big challenge for such experiment. In addition to the low threshold, the target of the detection system should be of low mass. The WIMP is expected to interact rarely with the detector material and hence the expected event rate from WIMP is very low which needs the detector to be of low intrinsic background and operated as in a background free environment. Several experiments like Dark Side-50, CRESST-III, NEWS-G, SuperCDMS, CDEX are exploring the low mass WIMP search using different detector techniques.  In the spin dependent sector, the PICASSO experiment provides limits in the WIMP mass region of 2 $GeV/c^2$ and 4 $GeV/c^2$ with a cross-section of $10^{-37}$ $cm^2$ \cite{2017APh.90.85B}. WIMP-electron scattering is another interesting channel to explore the low mass WIMPs and one of such experiments, SENSEI puts constraints in the 0.5-5 $MeV/c^2$ mass range for the WIMP-electron scattering that produces electron recoil induced signal \cite{PhysRevLett.122.161801}. For the dark matter search experiment, both the cosmogenic and radiogenic backgrounds contribute to the event of the experiment. To reduce the cosmogenic backgrounds, the dark matter experiments are usually planned at the underground labs \cite{Aubin_2008}. The superheated emulsion detector (SED) that employs the superheated liquid shows an excellent insensitivity to the backgrounds like gamma-rays by controlling the ambient temperature and pressure of the detector which essentially controls the threshold energy of bubble nucleation. Superheated liquid as a detector is used for the dark matter search experiments by different groups for a long time \cite{2013NIMPA.729.182M,SETH201692}, either the droplets of superheated liquids are suspended in a gel, polymer matrix or the bulk superheated liquid is used as a detector. The event occurs in such detector when an energetic particle/radiation deposits the energy in the medium which can satisfy the bubble formation conditions. During the bubble nucleation process, an acoustic signal is generated that is sensed by the piezo-electric sensor or microphones. Different parameters can be constructed from the acoustic signal and its Fast Fourier Transform which can be used to distinguish the different types of events in the detector \cite{PhysRevD.100.022001,PhysRevD.89.072013}. In the present work, the InDEx (Indian Dark matter search Experiment) is carried out at the 555m deep underground at the Jaduguda Underground Science Laboratory (JUSL), Jaduguda mine, Jharkhand, India. with $C_2H_2F_4$ (b.p. $-26.3^{\circ}C$) SED. This is the first experimental run of InDEx at the new underground laboratory. The earlier theoretical studies show that $C_2H_2F_4$ SED has the potentiality to prove the low mass WIMPs while operated at low thresholds \cite{PhysRevD.101.103005}. The background level at JUSL e.g. neutrons, gamma-rays, cosmic muons and the noise level at have been studies earlier \cite{SAHOO2021165450,2021NIMPA.99465083S,2022APh.13902700G}. In the present experiment, SED have been fabricated with $C_2H_2F_4$ active liquid and ran the experiment at JUSL for an exposure of 2.47 kg-days while operated the detector at 5.87 keV thresholds. The following sections explain the present experiment and analysis of the data collected at JUSL, the results followed by the discussion and finally the article ends with a conclusion section.

\section{Present Work}
The SED with the droplets of $C_2H_2F_4$ superheated liquid has been fabricated at the SINP laboratory and then transported the detectors to JUSL in a refrigerator. The droplets are suspended in a degassed gel matrix. First the gel matrix was prepared using aqua-sonic gel and glycerol and the matrix was degassed to remove the air bubbles which can act as the heterogeneous nucleation centers of bubble nucleation. The active liquid was transferred on the top of the gel in a high-pressure reactor and stirred it to make the droplets. The pressure was then reduced to the atmospheric pressure to make the superheated droplets in the gel matrix. The mixture was transferred to the borosil glass container which served as the SED. The detector was installed at JUSL at the room temperature of average $25.4^{\circ}C$. The piezo-electric acoustic sensor was coupled at the top of the detector touching the aqua-sonic gel. The output from the sensor was connected to the FPGA based data acquisition system \cite{2021NIMPA100965457S}. The effective run time of the detector was 48.6 days. The calibration was done with Am-Be (10 mCi) neutron source at the similar temperature as that of JUSL. The collected data were in the LabVIEW file format. The signals were analyzed using the Python code with different parameters constructed from the original signal and its Fast Fourier Transform. 
\section{Results and Discussion}
\begin{figure}[h]
	\centering
	\includegraphics[width=0.9\textwidth]{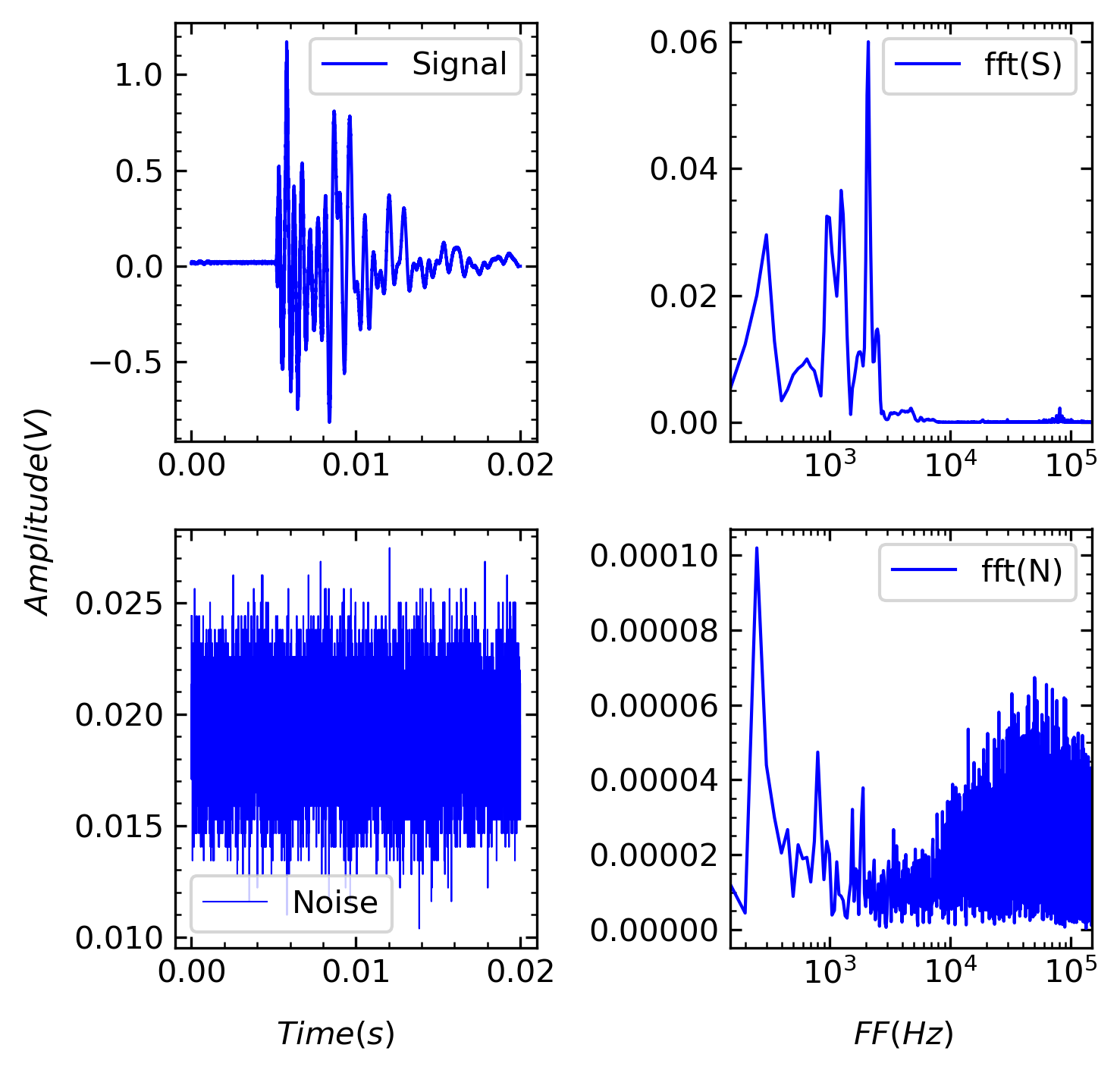}
	\caption{The bubble nucleation signal and the noise and its corresponding FFTs.}
	\label{fig:signal-noise-fft}
\end{figure}

\begin{figure}[h]
	\centering
	\includegraphics[width=0.9\textwidth]{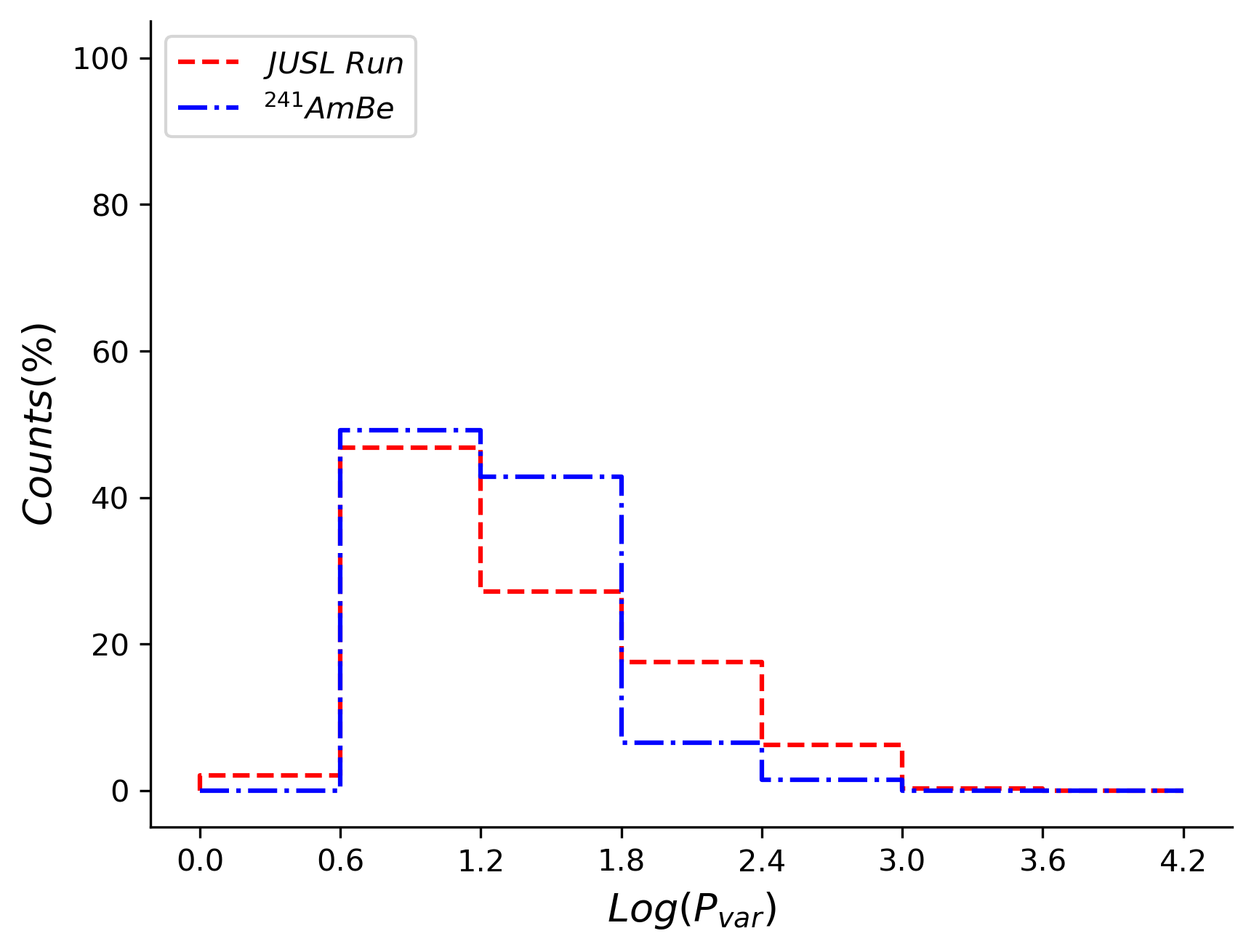}
	\caption{The variation of acoustic power for the JUSL run and calibration.}
	\label{fig:pvar-distribution}
\end{figure}

The noises and bubble nucleation signals at JUSL is shown in Fig.[\ref{fig:signal-noise-fft}]. The parameters Pvar is defined as the acoustic power which is obtained by summing the square of the amplitude of the signal over the duration of each signal which is shown in Fig.[\ref{fig:pvar-distribution}]. It is observed that the noises are well separated from the signal and in the analysis event by event separation of the noise has also been done by eye selection \cite{2022NIMPA102566186A}. It was observed earlier with condenser microphone that most of the noises at JUSL are in the audio-frequency range \cite{SAHOO2021165450}. The Fig.[\ref{fig:pvar-distribution}] shows that the signals from JUSL run are in the range of neutron induced signals of the calibration run. The main contributing background at 5.87 keV thresholds at JUSL is the neutrons. The estimated count rate for neutrons is found to be below the present observed count rate at JUSL.

\begin{figure}[h]
	\centering
	\includegraphics[width=0.9\textwidth]{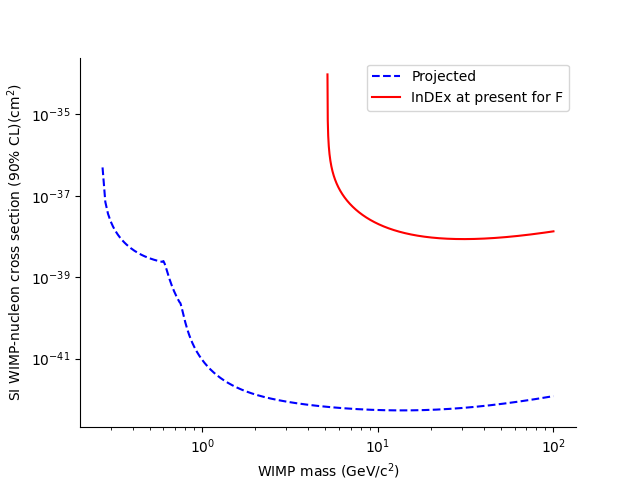}
	\caption{The result from InDEx run at JUSL for the Fluorine (red line) and the projected sensitivity of InDEx at 55$^{\circ}C$ for 100 kg-days of exposure under zero background consideration.}
	\label{fig:run1-sensitivity-projected}
\end{figure}

The estimated value of the upper limit at 90\% C.L. of the WIMP-nucleon cross-section for the spin-independent (SI) interaction with fluorine nuclei is shown in Fig.[\ref{fig:run1-sensitivity-projected}]. The efficiency for C and F is taken from the earlier published results \cite{2013NIMPA.729.182M,PhysRevD.101.103005}. For the estimation of cross-section, the methodology as explained in Ref \cite{PhysRevD.101.103005,1996APh.6.87L,2010JCAP.09.020C} is considered. The future projected sensitivity of InDEx for 100 kg-days at 0.19 keV thresholds have also been shown in Fig.[\ref{fig:run1-sensitivity-projected}] for the zero-background consideration. It is calculated earlier that the hydrogen nuclei in the target becomes sensitive for bubble nucleation at and below 0.19 keV thresholds which corresponds to the 55$^{\circ}C$ of the operating temperature \cite{PhysRevD.101.103005}. The upper limit of cross section appears at 30.67 $GeV/c^2$ for the SI-interaction with fluorine as $[7.939\pm(0.375)_{statistical}(^{+1.386}_{-0.909})_{systematic}]\times10^{-39}$ $cm^2$.

\section{Conclusion}
The superheated emulsion detector with $C_2H_2F_4$ active liquid has been fabricated at SINP lab and installed at JUSL along with the fabricated FPGA based data acquisition system. The experiment ran for 2.47 kg-days of exposure at a threshold of 5.87 keV. The upper limit on the spin-independent cross section for the fluorine nuclei have been presented with the most sensitivity at a WIMP mass of 30.67 $GeV/c^2$. The InDEx is sensitive from 4.44 $GeV/c^2$ WIMP mass for the Carbon nuclei. The result from the 1st run of InDEx at JUSL is presented in this article and subsequently InDEx will venture the low mass region with larger exposure and lower thresholds.
%
%

\bibliographystyle{unsrtnat}
\bibliography{references.bib}

\begin{thebibliography}{17}
\providecommand{\natexlab}[1]{#1}
\providecommand{\url}[1]{\texttt{#1}}
\expandafter\ifx\csname urlstyle\endcsname\relax
  \providecommand{\doi}[1]{doi: #1}\else
  \providecommand{\doi}{doi: \begingroup \urlstyle{rm}\Url}\fi

\bibitem[{Aalbers} et~al.(2024)]{2024arXiv241017036A}
J.~{Aalbers} et~al.
\newblock {Dark Matter Search Results from 4.2 Tonne-Years of Exposure of the
  LUX-ZEPLIN (LZ) Experiment}.
\newblock \emph{arXiv e-prints}, art. arXiv:2410.17036, October 2024.
\newblock \doi{10.48550/arXiv.2410.17036}.

\bibitem[Schumann(2019)]{Schumann_2019}
Marc Schumann.
\newblock Direct detection of wimp dark matter: concepts and status.
\newblock \emph{Journal of Physics G: Nuclear and Particle Physics},
  46\penalty0 (10):\penalty0 103003, aug 2019.
\newblock \doi{10.1088/1361-6471/ab2ea5}.

\bibitem[{Behnke} et~al.(2017)]{2017APh.90.85B}
E.~{Behnke} et~al.
\newblock {Final results of the PICASSO dark matter search experiment}.
\newblock \emph{Astroparticle Physics}, 90:\penalty0 85--92, April 2017.
\newblock \doi{10.1016/j.astropartphys.2017.02.005}.

\bibitem[Abramoff et~al.(2019)]{PhysRevLett.122.161801}
Orr Abramoff et~al.
\newblock Sensei: Direct-detection constraints on sub-gev dark matter from a
  shallow underground run using a prototype skipper ccd.
\newblock \emph{Phys. Rev. Lett.}, 122:\penalty0 161801, Apr 2019.
\newblock \doi{10.1103/PhysRevLett.122.161801}.

\bibitem[Aubin et~al.(2008)]{Aubin_2008}
F~Aubin et~al.
\newblock Discrimination of nuclear recoils from alpha particles with
  superheated liquids.
\newblock \emph{New Journal of Physics}, 10\penalty0 (10):\penalty0 103017, oct
  2008.
\newblock \doi{10.1088/1367-2630/10/10/103017}.

\bibitem[{Mondal} et~al.(2013){Mondal}, {Seth}, {Das}, and
  {Bhattacharjee}]{2013NIMPA.729.182M}
Prasanna~Kumar {Mondal}, Susnata {Seth}, Mala {Das}, and Pijushpani
  {Bhattacharjee}.
\newblock {Study of low frequency acoustic signals from superheated droplet
  detector}.
\newblock \emph{Nuclear Instruments and Methods in Physics Research A},
  729:\penalty0 182--187, November 2013.
\newblock \doi{10.1016/j.nima.2013.07.027}.

\bibitem[Seth and Das(2016)]{SETH201692}
Susnata Seth and Mala Das.
\newblock Radiation linear energy transfer and drop size dependence of the low
  frequency signal from tiny superheated droplets.
\newblock \emph{Nuclear Instruments and Methods in Physics Research Section A:
  Accelerators, Spectrometers, Detectors and Associated Equipment},
  837:\penalty0 92--98, 2016.
\newblock ISSN 0168-9002.
\newblock \doi{https://doi.org/10.1016/j.nima.2016.08.058}.

\bibitem[Amole et~al.(2019)]{PhysRevD.100.022001}
C.~Amole et~al.
\newblock Dark matter search results from the complete exposure of the pico-60
  ${\mathrm{c}}_{3}{\mathrm{f}}_{8}$ bubble chamber.
\newblock \emph{Phys. Rev. D}, 100:\penalty0 022001, Jul 2019.
\newblock \doi{10.1103/PhysRevD.100.022001}.

\bibitem[Felizardo et~al.(2014)]{PhysRevD.89.072013}
M.~Felizardo et~al.
\newblock The simple phase ii dark matter search.
\newblock \emph{Phys. Rev. D}, 89:\penalty0 072013, Apr 2014.
\newblock \doi{10.1103/PhysRevD.89.072013}.

\bibitem[Seth et~al.(2020)Seth, Sahoo, Bhattacharjee, and
  Das]{PhysRevD.101.103005}
Susnata Seth, Sunita Sahoo, Pijushpani Bhattacharjee, and Mala Das.
\newblock Probing low-mass wimp candidates of dark matter with
  tetrafluoroethane superheated liquid detectors.
\newblock \emph{Phys. Rev. D}, 101:\penalty0 103005, May 2020.
\newblock \doi{10.1103/PhysRevD.101.103005}.

\bibitem[Sahoo et~al.(2021)Sahoo, Ali, Das, Biswas, Pallav, and
  Basu]{SAHOO2021165450}
Sunita Sahoo, Suraj Ali, Mala Das, Nilanjan Biswas, Piyush Pallav, and Jisnu
  Basu.
\newblock The background study at 555 m deep underground with superheated
  emulsion detector.
\newblock \emph{Nuclear Instruments and Methods in Physics Research Section A:
  Accelerators, Spectrometers, Detectors and Associated Equipment},
  1008:\penalty0 165450, 2021.
\newblock ISSN 0168-9002.
\newblock \doi{https://doi.org/10.1016/j.nima.2021.165450}.

\bibitem[{Sharan} et~al.(2021){Sharan}, {Singaraju}, {Sinha}, {Ghosh}, and
  {Jha}]{2021NIMPA.99465083S}
Manoj~K. {Sharan}, Ram~Narayan {Singaraju}, Tinku {Sinha}, Tanmoy {Ghosh}, and
  V.~N. {Jha}.
\newblock {Measurement of cosmic-ray muon flux in the underground laboratory at
  UCIL, India, using plastic scintillators and SiPM}.
\newblock \emph{Nuclear Instruments and Methods in Physics Research A},
  994:\penalty0 165083, April 2021.
\newblock \doi{10.1016/j.nima.2021.165083}.

\bibitem[{Ghosh} et~al.(2022){Ghosh}, {Dutta}, {Mondal}, and
  {Saha}]{2022APh.13902700G}
Sayan {Ghosh}, Shubham {Dutta}, Naba~Kumar {Mondal}, and Satyajit {Saha}.
\newblock {Measurements of gamma ray, cosmic muon and residual neutron
  background fluxes for rare event search experiments at an underground
  laboratory}.
\newblock \emph{Astroparticle Physics}, 139:\penalty0 102700, June 2022.
\newblock \doi{10.1016/j.astropartphys.2022.102700}.

\bibitem[{Sahoo} et~al.(2021){Sahoo}, {Chaddha}, {Sahoo}, {Biswas}, {Roy},
  {Das}, and {Pal}]{2021NIMPA100965457S}
Shantonu {Sahoo}, Niraj {Chaddha}, Sunita {Sahoo}, Nilanjan {Biswas}, Anindya
  {Roy}, Mala {Das}, and Sarbajit {Pal}.
\newblock {FPGA-based multi-channel data acquisition system for Superheated
  Emulsion Detectors}.
\newblock \emph{Nuclear Instruments and Methods in Physics Research A},
  1009:\penalty0 165457, September 2021.
\newblock \doi{10.1016/j.nima.2021.165457}.

\bibitem[{Ali} and {Das}(2022)]{2022NIMPA102566186A}
Suraj {Ali} and Mala {Das}.
\newblock {Discrimination of neutron and gamma ray induced nucleation events at
  high frequency in R134a superheated emulsion}.
\newblock \emph{Nuclear Instruments and Methods in Physics Research A},
  1025:\penalty0 166186, February 2022.
\newblock \doi{10.1016/j.nima.2021.166186}.

\bibitem[{Lewin} and {Smith}(1996)]{1996APh.6.87L}
J.~D. {Lewin} and P.~F. {Smith}.
\newblock {Review of mathematics, numerical factors, and corrections for dark
  matter experiments based on elastic nuclear recoil}.
\newblock \emph{Astroparticle Physics}, 6\penalty0 (1):\penalty0 87--112,
  December 1996.
\newblock \doi{10.1016/S0927-6505(96)00047-3}.

\bibitem[{Chaudhury} et~al.(2010){Chaudhury}, {Bhattacharjee}, and
  {Cowsik}]{2010JCAP.09.020C}
Soumini {Chaudhury}, Pijushpani {Bhattacharjee}, and Ramanath {Cowsik}.
\newblock {Direct detection of WIMPs: implications of a self-consistent
  truncated isothermal model of the Milky Way's dark matter halo}.
\newblock \emph{jcap}, 2010\penalty0 (9):\penalty0 020, September 2010.
\newblock \doi{10.1088/1475-7516/2010/09/020}.

\end{thebibliography}

%


\end{document}